\documentclass[%
 prl, twocolumn,
 amsmath,amssymb,
 aps,
]{revtex4-2}

\usepackage{ifthen}
\newboolean{arxive}
\setboolean{arxive}{false} 
\newcommand{\notonarxive}[1]{\ifthenelse{\boolean{arxive}}{}{#1}} 
\newcommand{\onarxive}[1]{\ifthenelse{\boolean{arxive}}{#1}{}} 
\usepackage{xcolor}
\newcommand\todo[1]{?{\color{green}#1}?}
\newcommand\gtf[1]{{\color{orange}#1}}
\newcommand\mae[1]{{\bf \color{olive}#1}}
\newcommand\maeb[1]{\mae{[#1]}}
\newcommand\jl[1]{{\color{red}#1}}
\newcommand\sbd[1]{{\color{blue}#1}}
\onarxive{
\renewcommand\todo[1]{}
\renewcommand\gtf[1]{}
\renewcommand\mae[1]{#1}
\renewcommand\maeb[1]{}

\renewcommand\jl[1]{}
\renewcommand\sbd[1]{}
}

\usepackage{graphicx}
\usepackage{dcolumn}
\usepackage{bm}
\usepackage{xcolor}
\usepackage{xr}
\usepackage{xspace}
\usepackage{enumitem}
\usepackage{accents}

\DeclareMathAlphabet\mathbfcal{OMS}{cmsy}{b}{n}

\usepackage{color}
\usepackage{mathtools}
\usepackage{siunitx}
\usepackage{comment}
\usepackage{textcomp} 
\DeclareSIUnit\permille{\text{\textperthousand}}

\usepackage{subfigure}
\notonarxive{
}
\usepackage{lineno}
\newcommand{\etal}{\textit{et al.}\xspace}
\begin{document}

\title{Droplet absorption and spreading into thin layers of polymer hydrogels}

\author{Merlin A. Etzold$^1$, George T. Fortune$^{1}$, Julien R. Landel$^{2}$ and Stuart B. Dalziel$^1$
  \email{merlinaetzold@cantab.net}}
\affiliation{$^1$Department of Applied Mathematics and Theoretical 
Physics, Centre for Mathematical Sciences, University of Cambridge, Wilberforce Road, Cambridge CB3 0WA, 
United Kingdom \\
$^2$Department of Mathematics, Alan Turing Building, University of Manchester, Oxford Road, Manchester, M13 9PL, UK}%

\date{\today}
             
\begin{abstract} 
    From biological tissues to layers of paint, macroscopic non-porous materials with the capacity to swell when brought in contact with an appropriate solvent are ubiquitous. Here, we  study experimentally and theoretically one of the conceptually simplest of such systems, the swelling of a thin hydrogel layer by a single water drop. Using a  bespoke experimental setup, we observe fast absorption leading to a radially spreading axisymmetric blister. Employing a linear poroelastic framework and thin-layer scalings, we develop a non-linear one-dimensional diffusion equation for the evolution of the blister height profile, which agrees well with experimental observations.


\end{abstract}
\maketitle



High-swelling polymer systems, e.g. hydrogels (hydrophilic polymers in water systems), have received considerable attention recently \citep{Laftah11,Vervoort06}. Large volumetric strains, induced by swelling in response to changing environmental conditions, lead to a fascinating class of problems, where chemically-driven transport within the polymer drives large changes to the geometry of the hydrogel. These systems largely behave as hyperelastic incompressible solids on short time scales \cite{Yoon:2010}. Since the mechanical properties of hydrogels can be widely adjusted by changing the polymer chemistry and cross-linker density \citep{Tortorella21, Ozcelik16}, they have found wide-spread applications. 
The ability of external stimuli to trigger swelling that drives large deformations has been harnessed to drive microfluidic pumps \citep{Richter:2009,Kwon:2011} and has led to their use as micro-actuators for optical, flow control, sensor and microrobotics applications \citep{Ionov:2014, Porter:2007}. Understanding the interaction between swelling mechanics and microscopic driving forces is crucial for the design of these systems.

The excellent biocompatiblility of hydrogels has ensured their use in biomedical applications \citep{Hoffman:2012} for over sixty years since the pioneering development of the soft contact lens \cite{Wichterle60}. Since their physico-chemical characteristics are similar to many tissues \citep{Drury03,Hoffman12}, they have been used as tissue models in laboratory experiments (e.g. for decompression sickness \citep{Walsh:2017,Zhang:2020}). In tissue engineering, they are used as a matrix framework for the repair and regeneration of a wide range of tissues and organs \cite{Lee01,Tang20,Madhusudanan20}. Hydrogel wound dressings have been developed that maintain a moist healing environment whilst allowing gaseous exchange, promoting wound healing \citep{Tortorella21,Deng22,Stubbe21}. 

These systems, characterised by complex interactions between biological materials and hydrogels, require a thorough understanding of hydrogel swelling behaviour. An example of this interplay lies in the work of Cont \etal \cite{Cont20} who explore experimentally how \textit{Vibrio cholerae} biofilms deform both thin hydrogel layers and epithelial cell mono-layers attached to the surface of a soft extracellular matrix. The interactions between swelling and mechanical stresses cause the layers to buckle into the biofilm and often break up, thus allowing the biofilm to compromise the physiology of its host.

A range of fully three-dimensional poroelastic theories for hydrogel swelling have been proposed, 
deriving a constitutive equation for the stress tensor from the strain energy density function \cite{Yoon:2010,Doi:2009,Hong:2008,Chester:2010,Cai:2012,Bouklas:2012}. While linear theories assume this function to be quadratic in the strain \cite{Yoon:2010}, non-linear theories \citep{Doi:2009,Hong:2008,Chester:2010} construct it from thermodynamic models for polymer deformation and polymer--solvent mixing \citep{Cai:2012}. These models become equivalent in the limit of small deformation \citep{Bouklas:2012}. Nevertheless, 
most experimental studies have considered one-dimensional or quasi-one dimensional geometries only, e.g. the swelling of fibres \citep{vandewelde:2020}, spheres \citep{Tanaka:1979,Engelsberg:2013,Bertrand:2016} or flat sheets bound to surfaces being completely exposed to solvent \citep{Tanaka:1979}. More complicated problems are solved using finite element methods, e.g. the swelling caused by droplet trains \citep{Phadnis:2018}. However, considerable challenges remain, especially the resolution of enormous swelling gradients in high-swelling systems 
\citep{Yu:2020}.

Inspired by the flat geometries found in cell layers \citep{Cont20}, wound dressings \citep{Tortorella21}, hydrogel water harvesters \citep{Li:2018} and layers of paint \citep{Varady:2016}, in this letter we investigate experimentally and theoretically the absorption and spreading of a single water droplet into a thin hydrogel layer, which leads to blister formation and its slow outward spreading. This is also a model system for the absorption of hazardous substances into thin layers of paints and coatings \citep{Varady:2016}. 

\begin{figure}[!h]
    \centering
    \includegraphics[width=0.48\textwidth]{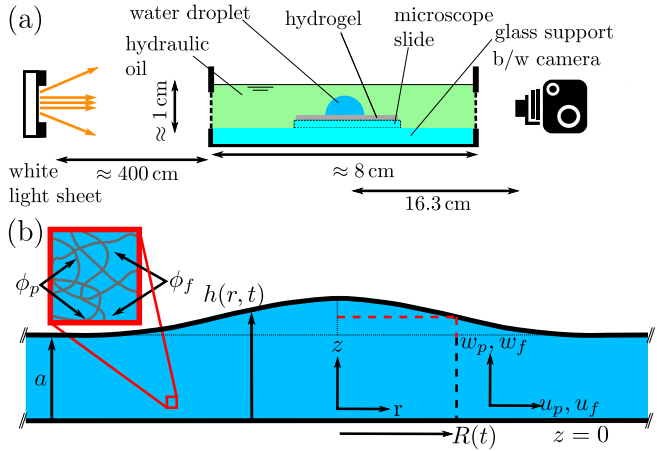}
    \caption{
    (a) Experimental apparatus. (b) Model hydrogel sheet. An axisymmetric blister (blue) with characteristic radius $R(t)$ spreads in an elastic layer fixed at $z = 0$ with a free surface at $z = h$, with undeformed height $a$. The pore-averaged velocities of the solid and fluid phases are  $\boldsymbol{u_p} = (u_p, \, w_p)$ and $\boldsymbol{u_f} = (u_f, \, w_f)$. For comparison with data we take $R(t)$ as the point at which the blister height decreases to half of its maximum value (red dashes). Inset: the hydrogel is a mixture of polymer (grey) and water (blue). 
    }  
    \label{fig:1}
\end{figure}

\newcommand{\photo}[1]{\ref{fig:1}(#1)}
We used commercially available hydrogel pads that were glued onto standard microscope slides and then placed in a mineral-oil filled bath (Fig. \ref{fig:1}). The hydrogel was then observed through windows in the bath against near-collimated background illumination.  The oil suppressed vapour phase transport due to evaporation and vapour absorption into the hydrogel. A single water droplet of volume $\{5,10,25,30,50, \SI{100}{\micro \litre}\}$ was then released  into the oil so that it sank to the hydrogel surface, forming a sessile droplet. Full experimental details are given in Supplementary Material (SM) \cite{suppmat} with representative profile data provided as Electronic Supplementary Information (ESI). 
\begin{figure*}
    \centering
    \includegraphics[width=\textwidth]{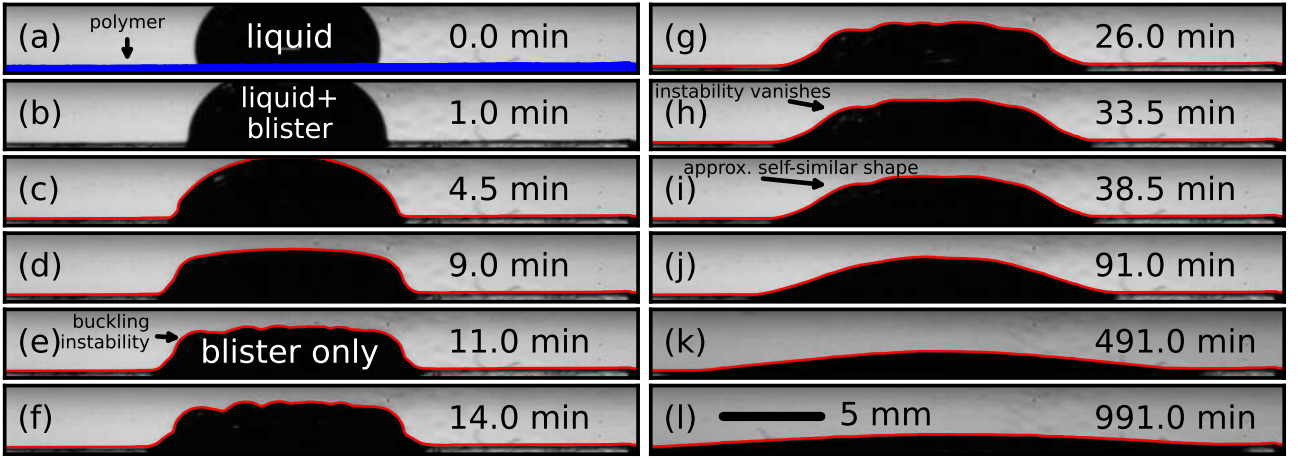}
    \caption{Sequence of photographs showing the absorption of a \SI{100}{\micro \litre} droplet into the hydrogel sheet. (a--d): States with liquid water remaining. (e--f): A surface instability becomes visible. (f--h): Transition towards the long-time spreading regime which is shown in (i--l). The visible hydrogel sheet is marked blue in (a). From (c) onward a red line marks the hydrogel--oil interface.}
    \label{fig:2}
\end{figure*}

Experimental footage for a \SI{100}{\micro \litre} droplet is shown in Fig. \ref{fig:2}. Due to refraction, the droplet and hydrogel appear dark. The upper hydrogel edge was located using automatic image analysis \cite{van2014scikit, harris2020}. The hydrogel swelling dynamics had three distinct phases. 
In the first phase, 
which lasted about \SI{10}{\minute} once the water had reached a sessile drop state on the hydrogel surface (after approximately \SI{30}{\second}), surface water and swollen hydrogel coexisted as the sessile droplet transformed into a swollen blister (Fig. \ref{fig:2}(a-c)). This transition was particularly apparent at the droplet edge, which became progressively steeper (Fig. \ref{fig:2}(c)) before becoming smooth again. 
In the second phase, once all the surface water had been absorbed, a surface instability appeared, forming a blister with a crinkled surface (Fig. \ref{fig:2}(e-g)). This buckling instability arises in hydrogels that have strong swelling gradients or are laterally confined \cite{Onuki:1988,Bouklas:2015}.
Supplementary experiments 
in air, where the water was removed at various stages of the first phase, found a thin swollen layer with the same surface instability whose wavelength increased with time. Hence, we conclude that this instability forms immediately after the droplet reaches the hydrogel surface, becoming visible once the droplet has been absorbed. In the third phase, which lasted until experiments were terminated once the blister front had reached the edge of the hydrogel, the buckling instability vanished as
 the blister assumed a self-similar shape resembling a non-linear diffusion process (Fig. \ref{fig:2}(h-l)). Further experimental aspects are reported in \citep{suppmat}.

Motivated by the self-similar profile observed in phase three of the experiments, we now develop a mathematical model for the swelling hydrogel layer using a poroelastic framework to obtain a non-linear diffusion equation describing its spreading dynamics. 
We consider a system in which the dynamics is
governed by the interplay between the motion of both polymer and water and the balance of elastic stresses through the polymer network. Fig. \ref{fig:1}(b) sketches the surface of a hydrogel layer of undeformed thickness $a$, fixed to a rigid horizontal boundary at $z = 0$. The gel is perturbed through the addition of a spherical water drop of diameter $d$, forming an axisymmetric blister of height profile $h(r,t)$ and characteristic radius $R(t)$.  
We examine the simplest hydrogel composition, namely a solution of polymer (volume fraction $\phi_p$, where $\phi_p = \phi_{0p}$ at the start of the experiment before the water drop is added) and water (modelled as a Newtonian fluid, volume fraction $1-\phi_p$). We denote the pore-averaged velocity and stress tensor of the solid and liquid phases by $\{ \boldsymbol{u_p} = (u_p, w_p) \, , \, \boldsymbol{\sigma_{p}} \}$ and $\{ \boldsymbol{u_f} = (u_f, w_f) \, , \, \boldsymbol{\sigma_{f}} \approx -p\boldsymbol{I} \}$ respectively, where $p$ is the pore pressure and $\bm{I}$ the identity tensor. Conservation of volume in the solid and fluid phases gives
\begin{equation}
\frac{\partial \phi_p}{\partial t} + \boldsymbol{\nabla \cdot} (\phi_p \boldsymbol{u_p}) = 0 \, , \,-\frac{\partial \phi_p}{\partial t} + \boldsymbol{\nabla \cdot} ((1 - \phi_p)\boldsymbol{u_f}) = 0. \label{eq:governing1and2} 
\end{equation}
Similarly, a momentum balance yields
\begin{equation}
\boldsymbol{\nabla \cdot \sigma} = \boldsymbol{\nabla}p, \label{eq:terzaghi}
\end{equation}
where $\boldsymbol{\sigma}$ obeys the elastic constitutive law $\boldsymbol{\sigma} = \boldsymbol{\sigma}(\boldsymbol{\nabla \xi})$ and $\boldsymbol{\xi} = (\xi, \zeta)$, is the 
displacement of the medium away from a reference state. This deformation relates to the  velocity of the polymer phase through $\boldsymbol{u_p} = \left(\partial_t + \boldsymbol{u_p \cdot \nabla}\right)\boldsymbol{\xi}$. Here, we consider the case where $\boldsymbol{\sigma}$ obeys the
linear constitutive law
\begin{equation}
    \boldsymbol{\sigma} = \left(K - \frac{2G}{3}\right)(\boldsymbol{\nabla \cdot \xi})\boldsymbol{I} 
    + G(\boldsymbol{\nabla \xi} + \boldsymbol{\nabla \xi}^T), \label{eq:governing3}
\end{equation}
 with $K$ and $G$  the osmotic and shear moduli of the hydrogel respectively, assumed constant \cite{Doi:2009,Etzold:2021}. We close the system of equations \eqref{eq:governing1and2}--\eqref{eq:governing3} by invoking Darcy's law for fluid motion within the matrix, 
\begin{equation}
(1 - \phi_p)(\boldsymbol{u_p} - \boldsymbol{u_f}) = \frac{\kappa}{\mu_f} \boldsymbol{\nabla}p. \label{eq:governing4}
\end{equation}
where the hydrogel sheet permeability $\kappa$ satisfies
\begin{equation}
    \kappa = \kappa_0 \left(\frac{\phi_{0p}}{\phi_p}\right)^{\beta}. \label{eq:finalpermeability}
\end{equation}
Here, $\kappa_0$ is the characteristic permeability scale while $\beta$ is a parameter to be determined by fitting to experimental data (see \cite{suppmat} for further details). We exploit a lubrication approximation, namely that the characteristic radial length scale, the initial radius of the blister $R_0 = R(t = 0)$, is much greater than the characteristic vertical length scale, the initial height $H_0 = h(r = 0, t = 0)$. We nondimensionalise the equations anisotropically using these length scales and define $\mathcal{P} = p/P_0, k = \kappa/\kappa_0$, such that
\begin{align}
    \rho &= \frac{r}{R_{0}}, \ \ \tau = \frac{U_0t}{R_0}, \ \ {\mathcal{R}} = \frac{R(t)}{R_{0}}, \nonumber \\
   \ \ {\cal H} &= \frac{h(r,t)}{H_{0}}, \ \ \mathcal{A} = \frac{a}{H_{0}} \ \ \mathcal{V} = \frac{d^3}{12 R_0^2 H_0}, \nonumber
 \end{align}
where the characteristic velocity scale $U_0 = \kappa_0(K+4G/3)/(R_0\mu_f)$ arises from the pressure gradients induced by the elastic stresses inherent in the polymer matrix. Keeping only leading-order terms in $\epsilon=H_0/R_0$ (see \cite{suppmat}), the model admits a vertically independent solution for the polymer volume fraction $\phi_{p}$ with the corresponding elastic deformation vector $\boldsymbol{\xi}$ satisfying
\begin{equation}
    \phi_p = \frac{\mathcal{A}}{\mathcal{H}}\phi_{0p}, \quad \boldsymbol{\xi} = z \left( 1 - \frac{\mathcal{A}}{\mathcal{H}} \right) \boldsymbol{e_{z}}. \label{eq:phisolution}
\end{equation}
The deflection of the hydrogel surface $\overline{\mathcal{H}} = \mathcal{H} - \mathcal{A}$ satisfies a conservation law of the form $\partial {\overline{\mathcal{H}}}/\partial \tau = - \boldsymbol{\nabla \cdot {\mathcal{J}}_{\mathcal{H}}}$,
\begin{align}
    \frac{\partial \overline{\mathcal{H}}}{\partial \tau} &= \frac{1}{\rho}\frac{\partial}{\partial \rho}\left( \rho \left(\frac{\overline{\mathcal{H}}+\mathcal{A}}{\mathcal{A}}\right)^{\beta - 1}  \frac{\partial \mathcal{H}}{\partial \rho} \right), \label{eq:finalconservationlaw}
\end{align}
where the flux $\mathcal{J}_{\mathcal{H}} = - \left(1 + \overline{\mathcal{H}}/ \mathcal{A}\right)^{\beta - 1}\boldsymbol{\nabla}_{\rho} \mathcal{H}$ is a plug-flow Darcy law term, dominated by the radial pressure gradient, together with the volume conservation condition
 $\int^{\infty}_0 \rho \overline{H} \, d\rho = \mathcal{V}. \label{eq:finalvolumecondition}  $
For general $\mathcal{A}$, (\ref{eq:finalconservationlaw}) does not admit an analytic solution. Hence, we solve it numerically using the finite-element software package FEniCS (\cite{AlnaesBlechta2015a}, \cite{LoggMardalEtAl2012a}, further details in section IV. A. of \cite{suppmat}. This system admits self-similar solutions
\begin{subequations}
\begin{align}
\mathcal{H}_{s} - \mathcal{A} &= \frac{1}{\mathcal{R}_s^2} \exp{\left(-\frac{(1 - \mathcal{A})}{2\mathcal{V}}\frac{\rho^2}{\mathcal{R}_s^2}\right)}, \label{eq:smalldeformationselfsimilar} \\
\mathcal{H}_l - \mathcal{A} &= \frac{(1 - \mathcal{A})}{\mathcal{R}^2_l}\left( 1 - \frac{(1 - \mathcal{A})(\beta - 1)}{2\beta\mathcal{V}} \frac{\rho^2}{\mathcal{R}_l^2} \right)^{1/(\beta - 1)}, \label{eq:largedeformationselfsimilar}
\end{align}
\end{subequations}
in the small ( $\mathcal{H}_{s}$, $1 - \mathcal{A} \ll 1$) and the large ($\mathcal{H}_l$, $\mathcal{A} \ll 1$) deformation limit respectively, with 
\begin{subequations}
\begin{align}
\mathcal{R}_s &= \left( 1 + \frac{2\tau(1 - \mathcal{A})}{\mathcal{V}} \right)^{1/2},    \label{eq:smalldeformationssradius} \\
\mathcal{R}_l &= \left( 1 + \frac{2 \tau \left( 1 - \mathcal{A} \right)^{\beta} {\mathcal{A}}^{\beta - 1}}{\mathcal{V}} \right)^{1/(2\beta)}.    \label{eq:largedeformationssradius}
\end{align}
\end{subequations}
The evolution of the blister will approach these similarity solutions asymptotically at large time, even when the initial conditions do not match the asymptotic form. 
\begin{figure}[!h]
    \centering
    \includegraphics[width=\columnwidth,keepaspectratio=true]{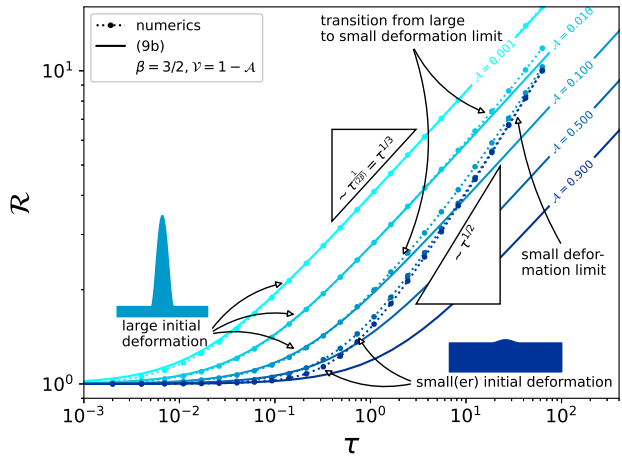}
    \caption{
    Spreading dynamics in thin hydrogels according to the proposed model with an arbitrary choice of $\beta = 3/2$, $\mathcal{V}=1-\mathcal{A}$ computed for numerical solutions of (\ref{eq:finalconservationlaw}) as a function of time, using  (\ref{eq:largedeformationselfsimilar}) as initial condition. Solid lines show the radius predicted by the self-similar solution in the large deformation limit (\ref{eq:largedeformationssradius}). 
    } 
    \label{fig:3}
\end{figure}
The time evolution of numerical solutions of (\ref{eq:finalconservationlaw}) for $\mathcal{R}$ is illustrated in Fig. \ref{fig:3} for $\beta=3/2$ and a range of values of $\mathcal{A}$ (darker blue denotes larger $\mathcal{A}$).
Dotted lines with circles denote 
numerical solutions of (\ref{eq:finalconservationlaw}) evolved from an initial condition of the form given in (\ref{eq:largedeformationselfsimilar}), where $\mathcal{R}$ is defined as the radius at which the blister height has decreased to half of the maximum height (see Fig. \ref{fig:1} and \cite{suppmat} (S2, S41)). 
Solid lines denote the corresponding large deformation similarity solutions (\ref{eq:largedeformationssradius}). For large deformations, ($\tau$ and $\mathcal{A}$ are small), the large deformation similarity solution (\ref{eq:largedeformationssradius}) agrees well with $\mathcal{R}$ obtained from the numerics, ($\mathcal{R} \sim \tau^{1/3}$). Since the maximum blister height $\mathrm{max}(\mathcal{H})$ decreases as time passes, the system eventually reaches the small deformation region with the corresponding similarity solution given for $\mathcal{R}$ in (\ref{eq:smalldeformationssradius}), ($\mathcal{R} \sim \tau^{1/2}$). 

Experimentally, $0.43<\mathcal{A}<0.68$ lies in the intermediate regime. This, together with the necessary resolution of a virtual origin problem, precludes a comprehensive direct quantitative test of the similarity solutions 
(which describe the behaviour of the blister after an initial adjustment period) against our experimental data. Instead, we solve (\ref{eq:finalconservationlaw}) numerically, using an experimentally observed initial condition obtained at $t=t_0$, with $t_0$ selected separately for each experiment to ensure both that the model assumptions were justified and that any buckling instability had decayed (see \cite{suppmat}).

The numerical solutions predict the evolution of the blister as a function of dimensionless time $\tau$ and $\beta$. For each experiment, the poroelastic time scale  $\tau_0 = \mu_f R_0^2 / \kappa_0 (K + 4G/3)$ (where $R_0$ differs for each experiment) is determined by globally fitting $\Omega = \tau_0/R_0^2 = \mu/(\kappa_0(K+4G/3))$, since $K$, $G$ and $\kappa_0$ and $\beta$ are unknown material parameters. 

We found $\Omega$ = \SI{6.72e9}{\second \per \metre \squared} and $\beta=2.25\pm0.083$ through a global fit of $\mathrm{max}(\mathcal{H})$ to all data sets \citep{suppmat}, leading to  $\tau_0 = 4.14 -\SI{36.79}{\hour}$ (the initial radii vary considerably, by almost a factor of 3, between the datasets). Whilst $\beta=2.25$ is considerably larger than the suggestion of $2/3$ made by \cite{Etzold:2021}, it is compatible with many previous experimental studies which found $1.5\leq \beta \leq 1.85$ \citep{Tokita:1991,Grattoni:2001,Bertrand:2016}. 
We found that the fit does not seem to be particularly sensitive to $\beta$; if we had chosen $\beta=1.83$, about the highest value reported in the literature \cite{Grattoni:2001}, $\Omega$ would have decreased slightly to \SI{5.82e9}{\second \per \metre \squared} (\cite{suppmat}).

In our model, we have assumed that the elastic deformation gradients in the $z$ direction are negligible. This assumption is reasonable since the characteristic poroelastic time scale $\tau_0$ is much greater than the time scale on which these gradients decay. Indeed, using experimental observations, we can relate these gradients to the buckling instability, which vanishes within approximately \SI{30}{\minute}. Since from above $\tau_0 = 4.14 -\SI{36.79}{\hour}$, this confirms empirically our assumption that the elastic deformation gradients in the $z$ direction are only important for the early time swelling dynamics, not captured by our model.

\begin{figure}
    \centering
    \includegraphics[width=\columnwidth, keepaspectratio=true]{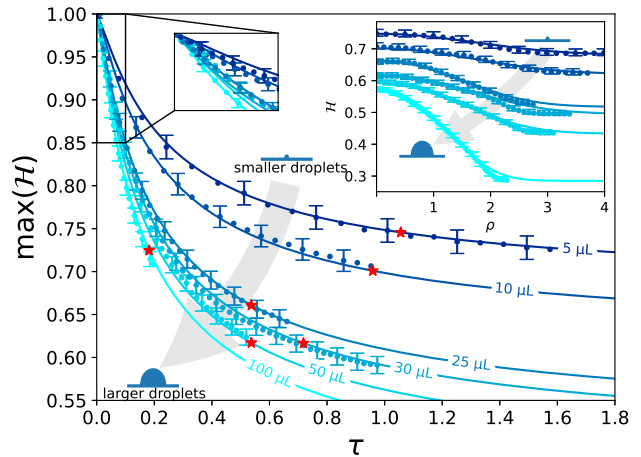}
    \caption{
    Comparison of experimentally obtained $\text{max}(\mathcal{H})$ (dots) to numerical solutions (solid lines). Inset: comparison of experimental profiles (dots), at the times indicated by the red stars, to numerical profiles (solid lines). 
    Error bars represent uncertainties \cite{suppmat}. 
    }
    \label{fig:4}
\end{figure}

Fig. \ref{fig:4} compares experimental data (dots) with numerical predictions (lines) generated using the model and the fitted parameters, taking initial conditions directly from the experimental data that they are fitted against (darker blue curves indicate larger initial droplets). The main figure explores how the maximum scaled blister height $\text{max}(\mathcal{H})$ evolves temporally for different droplet volumes. Very good agreement within the estimated uncertainty is found for all data sets for the full range of experimental data. The inset compares experimental height profiles of $\mathcal{H}(\rho)$ at a late time towards the end of each dataset (marked by a red star). Within the estimated uncertainty, the model and experiments agree well. However, for the larger droplets (lighter colours) the model slightly over predicts the front position. We attribute this to a combination of experimental difficulties in resolving the front of the swollen region and our choice of fitting method, since fitting $\mathrm{max}(\mathcal{H})$ ensures good agreement for $\rho=0$. In the Supplementary Material, we demonstrate that for $\mathcal{R}$, a quantity which is experimentally harder to access accurately, we see a similarly good agreement. Thus, after fitting against a single scalar observable ($\mathrm{max}(\mathcal{H})$) our model can quantitatively predict the evolution of all observable quantities over hours of observation time with only two fit parameters ($\beta$ and $\Omega$ where $1/\Omega$ is an effective poroelastic diffusivity). 
 
We believe these findings can aid the study of swelling phenomena in hydrogels and related polymers. Rather than constructing a model for the hydrogel using a fully non-linear constitutive equation \citep{Cai:2012}, we have demonstrated the efficacy of a linear poroelastic equation for $\bm{\sigma}$ combined with non-linear kinematic equations. Linear poroelastic models have a rigorous relationship with the aforementioned non-linear models \citep{Doi:2009,Bouklas:2012}. Due to the limited extensibility of the polymer chains, when constructing non-linear models for highly-swollen gels, it is necessary to use non-Gaussian statistics to model the changes in entropy due to mechanical stretching \citep{Chester:2010}. This hidden problem is avoided by linear models.  

In the lubrication limit, our fully three-dimensional model reduces to a single one-dimensional non-linear diffusion equation for the evolution of the non-dimensional height of the layer $\mathcal{H}$. This approach extends the geometries potentially amenable to analytical treatment beyond strictly one-dimensional geometries \citep{Yoon:2010,Engelsberg:2013,Bertrand:2016} and thus hopefully inspires future experimental work to be more ambitious than just one dimensional studies. Conversely, our dataset (with profile data provided as ESI) can be used to test alternative models for polymer swelling.

Our theoretical analysis does not cover the initial swelling phases (Fig. \ref{fig:2}(a-h)) which determine the initial conditions for the radial spreading problem. Previous modelling of quasi-one-dimensional hydrogels \citep{Etzold:2021}, spherical \citep{Bertrand:2016} hydrogels and similar strongly deforming poroelastic materials \citep{MacMinn:2015} suggested that the initial formation of an extremely swollen layer propagates into the material. Our experiments indicate that such a layer might have formed immediately after the droplet has made contact, possibly affecting the early-stage radial spreading rate. The formation of this layer is marked by a transient buckling instability; how this layer affects the absorption kinetics is unclear at this stage. Further work is needed to establish which parts of our theoretical frame work remain applicable in this case at early times.  

Importantly, our approach establishes experimental and theoretical connections between the polymer swelling literature and a large body of work considering spreading in thin porous layers \citep{Phillips09,Hewitt15,Nordbotten06,Rutqvist12}. This will accelerate further progress for complex swelling problems in slender geometries, where the polymer interacts with its environment in a complex fashion. Examples are the absorption of droplets on polymer materials \citep{Phadnis:2018}, mass transfer between a shear flow and swelling polymer surfaces \citep{Delavoipiere:2018} (see Landel \etal \citep{Landel:2016} for a non-swelling example) and biofilm growth on tissue \citep{Fortune21}.

\begin{acknowledgments}
The authors would like to thank the late Dr Henry McEvoy (Dstl, Porton Down) and  Professor M. Grae Worster (Cambridge) for proposing the problem.  The authors also thank Dr Steve Marriott (Dstl, Porton Down) for his help with the Karl-Fischer titration, MGW for his comments on the manuscript and Mr Paul Mitton and the late Mr Chris Summerfield for their advice and help during the experimental development. MAE, JRL and SBD acknowledge funding from Dstl under extramural research agreement DSTLX-1000138254. GTF acknowledges a Doctoral Training Fellowship from the Engineering and Physical Sciences Research Council. 
\end{acknowledgments}

\bibliography{hydrogel_references}

\end{document}


\title{Droplet absorption into thin layers of polymer: Supplementary Material}

\author{Merlin A. Etzold$^1$, George T. Fortune$^{1}$, Julien R. Landel$^{2}$ and Stuart B. Dalziel$^1$
  \email{merlinaetzold@cantab.net}}
\affiliation{$^1$Department of Applied Mathematics and Theoretical 
Physics, Centre for Mathematical Sciences, University of Cambridge, Wilberforce Road, Cambridge CB3 0WA, 
United Kingdom \\
$^2$Department of Mathematics, Alan Turing Building, University of Manchester, Oxford Road, Manchester, M13 9PL, UK}%

\date{\today}
             
\begin{abstract}
\end{abstract}
\maketitle
\section{Experimental}
\subsection{Supplementary Information on Methods and Materials}
We used commercially available medical grade hydrogel pads (Hydrogel Nipple Pads, Medela, Switzerland), which were manufactured as approximately $8\times\SI{8}{\centi \metre \squared}$ sheets. The hydrogel pads appear as rubbery sheets with a slightly tacky surface. After prolonged exposure to the atmosphere (circa 25-\SI{40}{\percent} relative humidity, circa \SI{22}{\celsius}), their initial thickness was determined with a micrometer as $a_0=\SI{1.13}{\milli \metre}\pm\SI{0.01}{\milli \metre}$. One side of a sheet was affixed to a thin  plastic sheet (impermeable to water) of approximate thickness \SI{0.04}{\milli \metre}. The other side was initially protected by an easily removable plastic film. The hydrogel sheets incorporate a thin gauze during their manufacture, probably to provide mechanical coherence and strength in the plane of the sheet. This gauze did not appear to noticeably affect the swelling of the sheets. In our experiments, these larger sheets were cut into approximately $25\times\SI{25}{\milli \metre \squared}$ coupons. The coupons were then glued (via the thin impermeable plastic film) onto $76\times\SI{26}{\milli \metre \squared}$ microscope slides (Menzel, Germany) with UV-cured Norland Optical Adhesive NOA 68 optical glue  (Thorlabs, USA).

Utilising Karl-Fischer titration (KFT), the water content of a typical hydrogel pad that had equilibrated with the laboratory air was measured to be \SI{22}{\percent}. Polymer swelling is driven by entropic and pairwise interactions between solvent and polymer molecules and charged groups if these are present \citep{FloryBook}. We tested for the presence of ionic groups by swelling tests in solvents which are less able to support the dissociation of the counter-ion.  We found that isopropanol and ethanol did not swell the hydrogel perceptively.  When placed in saturated sodium chloride solutions the swelling was drastically reduced. We take these observations as evidence for the presence of charged groups within the hydrogel. 

\subsection{Suppression of Vapour Transport}
In preliminary experiments, water droplets were placed onto mounted hydrogel that was otherwise exposed to air but enclosed into a sealed cell to prevent evaporation into the open atmosphere (the imaging system was similar to the present setup shown in the main text in Fig. 1(a)). The edges of the hydrogel sheet were observed to swell perceptively, with the swelling beginning immediately after droplet placement. This was attributed to evaporation from the strongly swollen regions (the regions in the vicinity of the droplet) leading to humidity gradients in the sealed cell and thus reabsorption by the drier regions of the hydrogel at the edge of the sheet. This was confirmed by experiments in which water droplets were placed next to the hydrogel on separate microscope slides within the sealed cell. Swelling was still observed which could have only have come from vapour phase transport.

To eliminate vapour transport experimentally, the hydrogel was immersed in a water-immiscible oil (CASTROL HySpin AWS 32). Since even apparently immiscible liquids are sparingly soluble into each other, we confirmed the water content of the hydraulic oil to be \SI{6700}{ppm} after prolonged exposure to the laboratory air using KFT. We also conducted a KFT with oil that had been equilibrated with a small amount of water, indicating an estimate for the equilibrium water content of the oil to be up to \SI{1.1}{\percent}. However, this might be an overestimation due to emulsion formation during shipping of the oil-water mixture to the place of analysis. To ensure that the oil and the hydrogel were in equilibrium, they were both exposed to the same atmosphere before the experiment for at least \SI{2}{\day}. The oil was not changed between experiments and thus was always exposed to the atmosphere. Furthermore, the hydrogel sheets were allowed to equilibrate for at least 1--\SI{3}{\hour} within the oil before droplet placement.

\subsection{Imaging and Data Analysis}
\subsubsection{Imaging System}
A monochrome camera (SP-5000M-CXP2, Jai, Denmark) with a variable magnification telecentric lens (TEV0305, VS Technology Corporation, Japan, utilising 0.4$\times$ magnification) aligned with the plane of the hydrogel sheet, recorded the evolution of our samples. Background lighting was provided by a white LED light sheet (MiniSun Light Pad 17031), masked with black tape and placed approximately \SI{4}{\metre} away from the hydrogel to ensure approximately collimated illumination (see Fig. 1(a) in the main text for a schematic of this setup). 

Digiflow was used for camera control \citep{Dalziel17} and captured images were analysed using a Canny edge detector from the Scikit Image Library \citep{van2014scikit} in Python. Suitable parameters for the edge detector (radius of Gaussian filter, upper and lower gradient thresholds for edge detection) were manually set for each dataset using a custom interactive interface. For each dataset, we extracted the position of the upper edge of the unswollen hydrogel before the droplet was placed (see Fig. 1(a) in the main text) and the upper edge of hydrogel and droplet once the droplet was placed (see Fig. 1(c-onwards) in the main text). The blister height (vertical displacement), defined as 
\begin{equation}
h_a = h(x,t)-a,
\end{equation}
with $0\leq x<x_{max}$ where $x_{max}$ is the right boundary of the cropped image, was then computed for each column of pixels. SciPy's Savitzky-Golay filter (window length $\SI{100}{pixel}= \SI{1.26}{\milli \metre}$, polynomial order 1) was used to smooth the raw $h_a$ data \citep{2020SciPy-NMeth}. The total height of the hydrogel $h(x,t)$ was thus determined by adding the initial thickness of the hydrogel sheet (determined by a micrometer). This procedure was adopted since the lower edge of the hydrogel was not visible in the images.

\subsubsection{Derived Quantities}
For a particular experimentally obtained profile $h(x,t)$, we found the centre position of the blister $x = r_0$ by fitting a Gaussian curve and taking $r_0$ to be the maximum of this curve. We define the corresponding characteristic radius of this blister profile $R(t)$ (indicated in Fig. 1(b) of the main text), as the radius at which the blister height $h_a(r+r_0,t)$ has decreased from the maximum height of the blister $\mathrm{max}(h_a(r+r_0,t))$ to half of this value. Mathematically, this can be written in algebraic form as
\begin{equation}
R = \left\{ r : h_a(r+r_0) = \frac{1}{2}\left(\underset{x}{\mathrm{max}}\left(h_a(x,t)\right)\right)  \right\}.\label{eq:practicalr}
\end{equation}
Due to the symmetry of the experimental profile about $r = r_0$, this leads to two values, $R_+$ and $R_-$ where $R_{-} \leq r_0 \leq R_{+}$, which we chose to average. We also compute the volume of the absorbed water by assuming cylindrical symmetry of the blister around $r_0$ using
\begin{equation}
    V = \pi \int_0^{x_{max}} (x-r_0)^2 h_a(x) dx,\label{eq:V}
\end{equation}
where the integral is evaluated over the full horizontal region of interest of the image (from $x = 0$ to $x = x_{max}$).
\subsubsection{Uncertainty Analysis}
\begin{figure}
    \centering
    \includegraphics[width=\columnwidth]{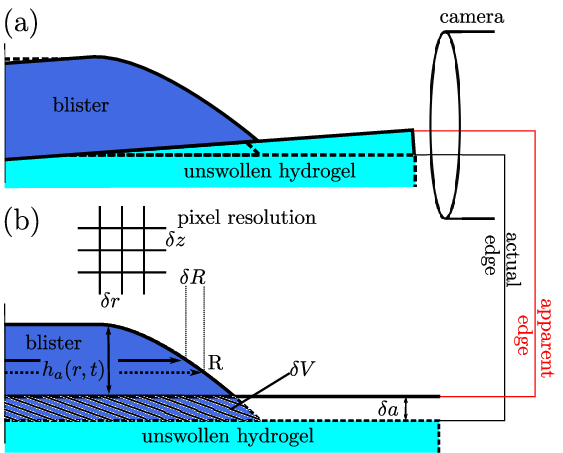}
    \caption{Illustration of the uncertainties inherent in the imaging. (a) Side view of the experimental configuration, showing how a misalignment (exaggerated for clarity) causes part of the droplet to be hidden from the camera by the edge of the hydrogel sheet. (b) Image (linear projection) of the droplet on the hydrogel sheet as seen by the camera. Dashed lines are not visible for the camera and give rise to an uncertainty in the position of the hydrogel surface $\delta a$, that creates a corresponding uncertainty $\delta R$ in the determination of $R$. The uncertainty due to finite pixel resolution in the radial and vertical directions are defined as $\delta r$ and $\delta z$ respectively. }
    \label{fig:imagingerrors}
\end{figure}

We assume our imaging system to be accurate within one pixel. Hence, at a pixel pitch of \SI{5}{\micro \metre} and a magnification of 0.4, this leads to imaging uncertainties $\delta r = \delta z = \SI{12.5}{\micro \metre}$. The particular configuration of the side view imaging leads to further uncertainties on the determination of $h_a$ due both to surface roughness (imperfections in the manufacturing process) and alignment errors between the camera and the sample, (illustrated in Fig. \ref{fig:imagingerrors}(a)). Fig. \ref{fig:imagingerrors}(b) shows how a small misalignment of angle $\delta \alpha$ causes some swelling to be hidden below the edge of the sample. This introduces an uncertainty in the position of the upper edge of the hydrogel $\delta a$. Assuming that the alignment between the camera and the sample is accurate within \SI{0.05}{\degree}, we find that for \SI{26}{\milli \metre} wide samples the edge position of the unswollen hydrogel appears a distance $\delta a^+=\SI{+22.7}{\micro \metre}$ too high. We thus estimate the uncertainty for the blister profile as $\delta h_a = (+\delta a, -\delta z) = (+22.7,-12.5)\si{\micro \metre}$. 

Fig. \ref{fig:imagingerrors}(b) illustrates the potentially significant effect that $\delta h_a$ may have on the uncertainties of $R$. We estimate these from the slope of $h_a$ at $R$ as
\begin{equation}
    \delta R \approx \left.\delta h\left/ \frac{\partial h_a}{\partial x}\right.\right|_{x=r_0+R}, \label{eq:dr}
\end{equation}
where the derivative is approximated for each frame using a fourth order central difference scheme. Our analysis (\ref{eq:dr}) shows that, if the slope of the blister surface decreases over time, the uncertainties grow over time. This makes the experimental determination of $R$ difficult at late stages when the blister becomes flat.

The uncertainty of the determination of $\delta h_a$ also effects the computation of $V$, (illustrated by the hatched region in Fig. \ref{fig:imagingerrors}(b)). We estimate the uncertainty $\delta V$ on the determination of the blister volume as
\begin{equation}
    \delta V \approx \pi \int_{x_{mn}}^{x_{mx}} (x-r_0) \delta h_a(x) \mathrm{d}x, \label{eq:volumerror}
\end{equation}
where the integration bounds $[x_{mn}, \, x_{mx}]$ are defined so that 
\begin{equation}
    \left\{x_{m}\leq x\leq x_{mx}: h_a(x-r_0)>\delta z \right\}.\label{eq:volumeintegrationbounds}
\end{equation}
This approximates the hatched region given in \ref{fig:imagingerrors}(b). We decided against extrapolating the exact edge position based on the slope since the slope of $h(x,t)$ becomes very small at the leading edge (indicated by both our theory and experimental data; see Fig. 2 in the main text at late times). Similar to the error analysis for $R$, the above analysis shows that accurately determining $V$ becomes progressively harder as the blister spreads, since $h_a$ decreases over time whilst the area covered by the blister, defined by (\ref{eq:volumeintegrationbounds}), increases with time.  We also note that the accurate determination of the volume relies on the assumption of cylindrical symmetry of the blister.

\subsubsection{Radius Raw Data}
\begin{figure*}
    \centering
    \includegraphics[width=\textwidth,keepaspectratio=true]{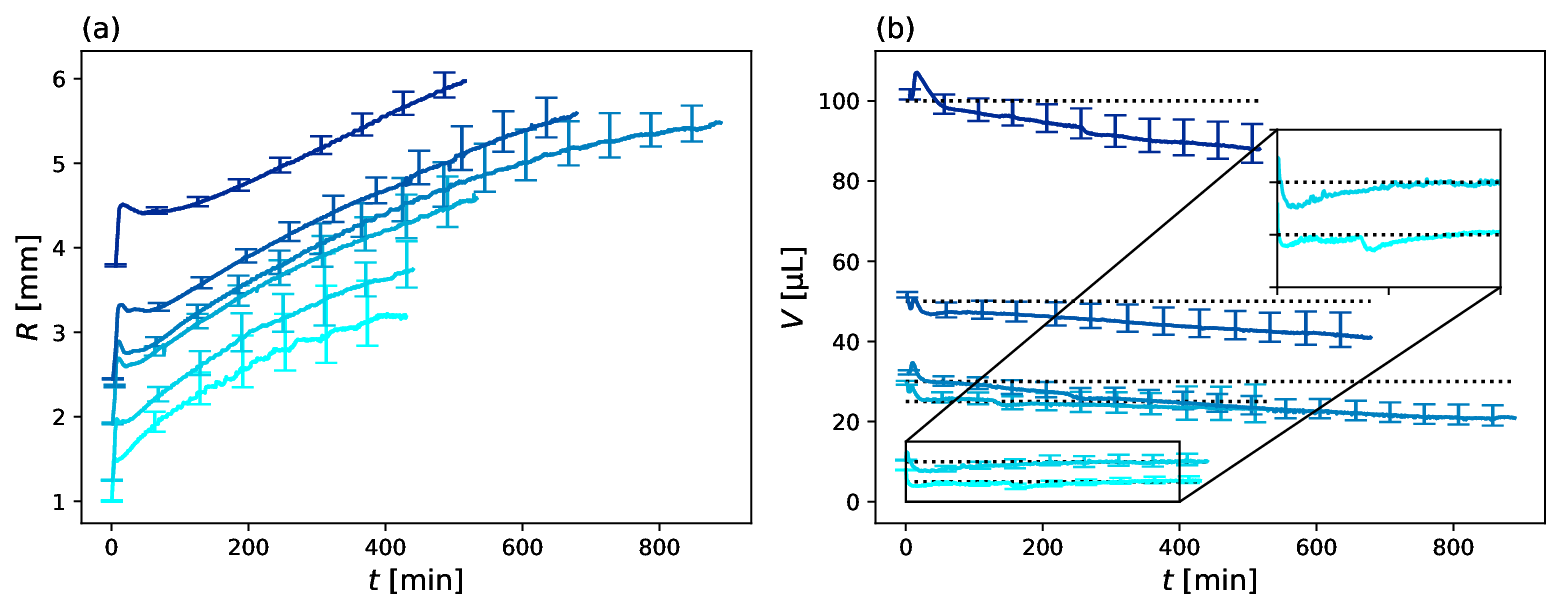}
    \caption{Raw data with corresponding error bars showing for a range of experiments how in (a) the radius of the swollen region and in (b) the apparent volume of the blister (found by direct integration) vary as functions of time. Darker blue curves denote experiments with larger water droplets.
    }
    \label{fig:consvolumeandradiusraw}
\end{figure*}

In Fig. \ref{fig:consvolumeandradiusraw}(a) we show dimensional data for the radius of the swollen region $R$ for the full duration of an experiment from the moment that the droplet makes contact with the hydrogel surface. Initially, $R$ sharply increases as the water droplet, having made contact, spreads slightly over the surface of the hydrogel (see Fig. 2(a) in the main text). Next, $R$ decreases slightly, as the morphology of the blister changes from being a relatively flat-capped shape, caused by the horizontal redistribution of water above the droplet as it is absorbed at a rate that is nearly constant over the entire droplet radius, towards a smoother more self-similar shape (see Fig. 2(b) to 2(h) in the main text). Finally, we reach a regime where the radius of the swollen region increases monotonically (see Fig. 2(i) to 2(l) in the main text). 
The error bars were determined as described above and increase with time due to the decreasing slope of the blister. This highlights the difficulty in determining $R$ experimentally for small droplets.

\subsection{Conservation of Volume}
Conservation of volume is an important assumption for our theoretical modelling. In this section we briefly investigate the validity of this assumption.

\subsubsection{Optically Determined Blister Volume}
In Fig. \ref{fig:consvolumeandradiusraw}(b), we report the blister volumes found by integrating the experimental height profile (\ref{eq:V}). Corresponding error bars represent our uncertainty estimation (\ref{eq:volumerror}). We find that for the smallest three droplets, (namely those with volume 5, 10 and \SI{25}{\micro \litre}), volume is largely conserved, in line with our assumptions. For the remaining larger droplets, we observe a continuous decrease in volume over the observed period. Whilst our uncertainty estimates capture the trend and magnitude of this decrease well, it cannot explain the magnitude of the decrease in volume for the two largest droplets. 

However, as noted in the error analysis above, accurately determining the blister volume becomes progressively difficult with increasing blister radius. Its accuracy relies strongly on the assumption of cylindrical symmetry. Given the instrumentation available at the time, cylindrical symmetry of the droplet settling on the hydrogel could only be verified by eye and thus might not have been perfect in some cases. We also observed that the bounds of the leading edge of the droplet (see \ref{eq:volumeintegrationbounds}) sometimes varied by up to \SI{10}{\percent}, indicating imperfect spherical symmetry. Surprisingly, the experiments with smaller droplets exhibit conservation of volume, whilst the larger droplets appeared to loose some volume. Smaller droplets experience stronger capillary forces and we would therefore expect them to settle more symmetrically on a surface compared to a big droplet.

We note that such effects are also visible in the early stages of the experiments, when the initial volume was overestimated. This is attributed to the initial creasing of the surface of the hydrogel. Only the silhouette of the external envelope is visible to the side-view camera, hiding the surface depressions associated with the creasing, and thus the creases would not be accounted for in the image processing integration step. 

Conservation of volume for the smaller droplets and the success of our error analysis to explain the trends seen in Fig. \ref{fig:consvolumeandradiusraw}(b) suggest that volume is nearly conserved. In particular, although we measure a significant volume loss, the majority of this loss can be attributed to measurement error rather than actual volume loss. However for completeness, we also estimated potential losses of water through the covering oil.

\subsubsection{Water Transport into or through the Oil}
Since there is a sparse solubility of water in the hydraulic oil (see above), we investigated the water losses due to transport into the oil phase. Here, we make an order of magnitude estimation for the volume loss from this  transport by considering the effect from diffusive mass flux in the $z$-direction. We assume that mass flux is due to diffusion only (no advective component) and that the oil is saturated with water at the interface to the hydrogel swollen region (estimated to be \SI{1.1}{\percent} water content which is an overestimation since the chemical potential of water in unsaturated hydrogel is less than of pure water \citet{Etzold:2021}) and is in equilibrium with the laboratory atmosphere far away from the hydrogel (estimated to be \SI{0.67}{\percent} water content). Hence, we estimate the difference in water volume concentration $\Delta c$ that drives water diffusion through the oil as
\begin{equation}
    \Delta c \approx \Delta \omega \rho_o,
\end{equation}
where the difference in water mass fraction that drives diffusion $\Delta \omega = \SI{0.43}{\percent}$, the density of mineral oil $\rho_o=\SI{833}{\kilo \gram \per \cubic \metre}$ and we have assumed both ideal mixing and that the water concentration in the oil is small. At a fixed radial distance $r = r'$, the concentration field $c = c(r',z,t,t')$ can be described using the similarity solution for the one-dimensional diffusion equation, (neglecting in-plane diffusion along the $(r,\theta)$ plane),
\begin{subequations}
\begin{equation}
c = \Delta c \, \mathrm{erfc}(z/\sqrt{4D(t-t')}),
\end{equation}
with corresponding diffusive water mass flux
\begin{equation}
j = \frac{\Delta c\sqrt{D}}{\sqrt{\pi (t-t')}}, 
\end{equation}
\end{subequations}
where $\mathrm{erfc}$ is the complementary error function, $D$ is the water diffusion coefficient and $t'$ is the time at which the swelling front reaches the point $r = r'$ ($r' = R(t')$). The total water volume lost for a swollen blister thus becomes
\begin{equation}
    \Delta V = \frac{1}{\rho_w} \int_0^{t}\left[ A_0 j(t') + \frac{\partial A}{\partial t}(t') \circledast j(t') \right] dt',
\end{equation}
where we have assumed that the blister has a circular base and surface area $A(t)$ with $A(0)=A_0$. The convolution integral denoted by $\circledast$ accounts for the fact that diffusion starts only when the hydrogel starts to swell. Assuming the radius to grow linearly with growth rate $\chi$, namely $A(t)=\pi (R_0+\chi t)^2$, this simplifies to become
\begin{align}
    \Delta V &= \frac{2\rho_o}{\rho_w} \sqrt{\pi D} \Delta \omega \left[R_0^2\sqrt{t}+\int_0^{t} \int_0^t \chi \frac{(R_0+\chi t'')}{\sqrt{t'-t''}} dt'' dt'\right] \nonumber \\
    &= \frac{2\rho_o}{\rho_w}\sqrt{\pi D} \Delta \omega \left[R_0^2 \sqrt{t}+\frac{4}{15}\chi t^{3/2}(5R_0+2\chi t)\right]. \label{eq:diffusivevolumelossprediction}
\end{align}
Utilising the experimental data for the largest blister (see Fig. 2 in main text and table \ref{tab:initialconditiondata}) 
with initial radius $R_0=\SI{5}{\milli \metre}$, approximating the radius of the hydrogel area covered by the blister as growing linearly with time yields an effective $\chi=\SI{0.5}{\milli\metre\per \hour}$ (this was roughly done based on imaging data (e.g. compare Figs. 2(c) and 2(k) in main text) rather than using $R$ as defined in (\ref{eq:practicalr}) since for flat blisters our $R$ underestimates the edge position). We found no water diffusion data for  the oil and established that such data is not widely available for mineral oils. However, given its relatively high viscosity of $\mu_o=\SI{32}{\centi St}$ at \SI{40}{\celsius}, the water diffusion coefficient in oil is likely to be an order of magnitude less than the self diffusion coefficient of bulk water, $D=\SI{3e-9}{\metre \squared\per \second}$. Furthermore,  Hilder and van den Tempel reported water diffusion coefficients in groundnut oil (50 times more viscous than water) and in kerosene (twice as viscous as water) as \SI{2.5e-10}{\metre \squared \per \second} and \SI{8.5e-10}{\metre \squared \per \second} respectively \citep{Hilder:1971}. Hence, taking the diffusion constants in groundnut oil and in kerosene as lower and upper bounds respectively for the diffusion constant of water in our hydraulic oil, (\ref{eq:diffusivevolumelossprediction}) predicts diffusive volume losses of $\{0.3-0.6 \}$, $\{ 1.2-2.2 \}$, and $\{2.7-5.0\}$ \si{\micro \litre} after one, five and ten hours respectively. As this is almost an order of magnitude too small to account for the volume loss found in Fig. \ref{fig:consvolumeandradiusraw}(b), we must seek an alternative explanation.

For completeness, a supplementary experiment was conducted where a \SI{2}{\micro \litre} droplet of water was placed directly onto a microscope slide that was immersed into an oil bath with no hydrogel present. This droplet reached a sessile state after \SI{25}{\hour}. Whilst the accuracy of our image analysis of this experiment was impacted from resolution issues in locating the diffuse edges of the droplet, we found that the water loss of the droplet was less than \SI{0.1}{\micro \litre} over a time period of approximately \SI{120}{\hour}. Its surface area remained approximately constant and was estimated to be \SI{4.6}{\milli \metre \squared}. Assuming a steady state mass loss from the droplet into the atmosphere through a \SI{1}{\centi \metre} deep layer of oil, this analysis yields a diffusion coefficient of 1-\SI{2e-10}{\metre \squared \per \second}. By approximating this curved diffusion problem using the equations above, assuming mass transport into an infinite oil bath yields an even lower diffusion coefficient of \SI{7e-11}{\metre \squared\per \second}. We note that these droplets remained in the oil bath for approximately one month over the summer without vanishing. 

\subsubsection{Conclusion}
It appears unlikely that, based on water diffusion coefficient from literature, the swollen hydrogel lost a significant amount of water into the oil during the duration of an experiment, (corroborated by a supplementary experiment testing the stability of a small water droplet in the same conditions). Also, the experimental observation that volume appears to be conserved for smaller droplets suggests that the extra volume loss apparent in Fig. \ref{fig:consvolumeandradiusraw}(b) for larger droplets beyond that attributed to measurement error is likely due to the initial blister profile not being completely axisymmetric, resulting in spreading that is not quite radial.

\section{Poroelastic Model} \label{poroelasticappendix}
\subsection{Modelling the Permeability of Hydrogel}
We model the dimensionless effective hydrogel permeability $k = \kappa / \kappa_0$ as an explicit dimensionless function of the polymer volume fraction $\phi_p$. Since the true structure of these materials at the pore scale is complicated and unknown for our particular sample, this can only be an approximation. In the porous media literature, many different models for the concentration-dependence of $\kappa$ have been used. In a geophysical context, Hewitt $\etal$ \cite{Hewitt15} for mathematical simplicity took the permeability of a shallow deformable porous layer as $k = 1$.  Modelling bacterial biofilms, Seminara $\etal$ \cite{Seminara12} set the permeability of a growing biofilm as $k = (1 - \phi)^2$ while Fortune $\etal$ \citep{Fortune21} constructed the system in such a way to not have to explicitly define $k$. When considering a spherical hydrogel, Betrand $\etal$ \citep{MacMinn:2016} proposed $k=(1-\phi_p)/\phi_p^\beta$. In the interest of simplicity we follow Tokita \& Tanaka  \citep{Tokita:1991} and Etzold $\etal$ \citep{Etzold:2021} by defining
\begin{equation}
    k = \left(\frac{\phi_{0p}}{\phi_p}\right)^{\beta}, \label{eq:finalpermeability}
\end{equation}
where $\beta$ is a free parameter chosen by matching to experimental data and $\phi_{0p}$ is a reference porosity. For a swelling cuboidal network $\beta = 2/3$ has been proposed by \cite{Etzold:2021}, but experiments have suggested values between $\beta=1.5$ \citep{Tokita:1991} and 1.85 \citep{Grattoni:2001}.
\subsection{Dimensionless shallow-layer scalings}
We scale radial and vertical lengths with the initial radius $R_0 = R(t = 0)$ and height $H_0 = h(r = 0, t = 0)$ of the blister respectively, (namely $\{ r, R \} \sim R_0$ and $\{ z, \, h \} \sim H_0$). Since the characteristic time scale $\tau_0$ for the system is the poroelastic time scale for pressure driven fluid flow in the horizontal direction, we scale $t \sim \tau_0$ where
\begin{equation}
    \tau_0 = \frac{\mu_f R_0^2}{(K+4G/3)\kappa_0}.\label{eq:poroelastictime}
\end{equation}
From equation (1) of the main text, we have $u_f \sim U_0 = R_0 / \tau_0$ and $\{ w_f, \, w_p \} \sim H_0 / \tau_0$. As horizontal and vertical elastic stresses in the hydrogel are of the same magnitude, $u_p \sim w_p \sim H_0 / \tau_0$. Since $\boldsymbol{u_s}$ is defined as the material derivative of $\boldsymbol{\xi}$, we have $\{ \zeta, \, \xi \} \sim H_0$. By definition $\kappa \sim \kappa_0$. Next, equation (4) of the main text implies that $\{ p, \, \tilde{p} \} \sim P_0 = K + 4G/3$. Finally $d \sim \left( 12 R_0^2 H_0 \right)^{1/3}$.
\subsection{Non-dimensional Governing Equations}
We now exploit a lubrication approximation, namely that the characteristic radial length scale $R_0$ is much greater than the characteristic vertical length scale $H_0$. We nondimensionalise the dimensional governing equations given in the main text anisotropically using the scalings given above, denoting the dimensionless form of a function $f$ by $f^{*}$ 
and setting
\begin{align}
    \rho &= \frac{r}{R_{0}}, \ \tau = \frac{t}{\tau_0}, \ {\cal H} = \frac{h(r,t)}{H_{0}}, \ {\mathcal{R}} = \frac{R(t)}{R_{0}}, \nonumber \\
    \mathcal{P} &= \frac{p}{P_0}, \ k = \frac{\kappa}{\kappa_0}, \ \mathcal{A} = \frac{a}{H_{0}}, \ \mathcal{V} = \frac{d^3}{12 R_0^2 H_0}. \nonumber
\end{align}
Keeping only leading-order terms in the aspect ratio $\epsilon = H_0/R_0\ll 1$, the system of equations (1)-(4) in the main text becomes
\begin{subequations}
\begin{align}
\frac{\partial \phi_p}{\partial \tau} &+ \frac{\partial}{\partial z^{*}}(\phi_p w_p^{*}) = \mathcal{O}\left(\epsilon \right), \label{eq:supgoverning1} \\
-\frac{\partial \phi_p}{\partial \tau} &+ \frac{\partial}{\partial z^{*}}\left( (1 - \phi_p) w_f^{*} \right) \nonumber \\
&+ \frac{1}{\rho}\frac{\partial}{\partial \rho}\left( \rho(1 - \phi_p)u_f^{*} \right) = 0, \label{eq:supgoverning2} \\
    w_f^{*} &- w_p^{*} = - \frac{k}{\epsilon^2(1 - \phi_p)} \, \frac{\partial \mathcal{P}}{\partial z^{*}}, \label{eq:supgoverning3} \\
    u_f &= - \frac{k}{(1 - \phi_p)} \, \frac{\partial \mathcal{P}}{\partial \rho} + \mathcal{O}\left( \epsilon \right), \label{eq:supgoverning4} \\
    \frac{\partial \mathcal{P}}{\partial \rho} &= \frac{1}{\epsilon}\left( \frac{G}{K + 4G/3} \right)\frac{\partial^2 \xi^{*}}{\partial {z^{*}}^2} \nonumber \\
    &+ \left( \frac{K + G/3}{K + 4G/3} \right)\frac{\partial^2 \zeta^{*}}{\partial \rho \partial z^{*}} + \mathcal{O}\left( \epsilon \right), \label{eq:supgoverning5} \\
    \frac{\partial \mathcal{P}}{\partial z^{*}} &= \frac{\partial^2 \zeta^{*}}{\partial {z^{*}}^2} \nonumber \\
    &+ \epsilon \left( \frac{K + G/3}{K + 4G/3} \right)\frac{1}{\rho}\frac{\partial}{\partial \rho}\left( \rho \frac{\partial \xi^{*}}{\partial z^{*}} \right) \nonumber \\
    &+ \mathcal{O}\left( \epsilon^2 \right), \label{eq:supgoverning6}
\end{align}
where
\begin{align}
    u_p^{*} &= \frac{\partial \xi^{*}}{\partial \tau} + w_p^{*} \frac{\partial \xi^{*}}{\partial z^{*}} + \mathcal{O}\left( \epsilon \right), \\
    w_p^{*} &= \left(1 - \frac{\partial \zeta^{*}}{\partial z^{*}}\right)^{-1} \frac{\partial \zeta^{*}}{\partial \tau} + \mathcal{O}\left( \epsilon \right). \label{eq:supgoverning7}
\end{align}
\end{subequations}
\subsection{Boundary Conditions}
\begin{subequations}
The bottom of the hydrogel layer is glued to an impermeable rigid glass slide, meaning
\begin{align}
    w_f^{*} &= 0 \mbox{ and } \xi^{*} = \zeta^{*} = 0 \Longrightarrow \nonumber \\
    u_p^{*} &= w_p^{*} = 0 \mbox{ at } z^{*} = 0. \label{eq:supbc1}
\end{align}
The kinematic boundary conditions at the upper surface $z^{*} = \mathcal{H}$ are
\begin{align}
    w_p^{*}\Big{|}_{\mathcal{H}} &= \frac{\partial \mathcal{H}}{\partial \tau} + \epsilon u_p^{*}\Big{|}_{\mathcal{H}} \frac{\partial \mathcal{H}}{\partial \rho}, \label{eq:supbc2a} \\
    w_f^{*}\Big{|}_{\mathcal{H}} &= \frac{\partial \mathcal{H}}{\partial \tau} + u_f^{*}\Big{|}_{\mathcal{H}} \frac{\partial \mathcal{H}}{\partial \rho}. \label{eq:supbc2b}
\end{align} \label{eq:supbc2}
Vertically integrating (\ref{eq:supgoverning1}) and then applying (\ref{eq:supbc2}) gives the polymer volume conservation condition
\begin{equation}
    \frac{\partial}{\partial \tau}\left( \int^{\mathcal{H}}_{0} \phi_p dz^{*} \right) = - \epsilon \left(\phi_p u_p^{*}\right)\Big{|}_{\mathcal{H}}\frac{\partial \mathcal{H}}{\partial \rho}. \label{eq:supbc3} 
\end{equation}
The stress-free boundary at $z^{*} = \mathcal{H}$ gives
\begin{align}
    (\sigma^{*})_{\rho z^{*}} \Big{|}_{\mathcal{H}} = 0 & \Longrightarrow \frac{\partial \xi^{*}}{\partial z^{*}}\Bigg{|}_{\mathcal{H}} = 0 + \mathcal{O}(\epsilon), \label{eq:supbc4} \\
    \left(\mathcal{P} - (\sigma^{*})_{z^{*}z^{*}} \right)\Big{|}_{\mathcal{H}^-} &= {\mathcal{P}\Big{|}_{{\mathcal H}^-} - \left.\frac{\partial \zeta^{*}}{\partial z^{*}}\right|_{H^-}} + \mathcal{O}\left( \epsilon \right) \nonumber \\
    &= \mathcal{P}\Big{|}_{\mathcal{H}^{+}} = \mathcal{P}_0. \label{eq:supbc5}
\end{align}  
Global water conservation requires
\begin{equation}
    \int^{\infty}_0 \rho \left({\mathcal{H}}-\mathcal{A} \right) \, d\rho = \mathcal{V}. \label{eq:supbc6}
\end{equation}
Finally, in the far field the polymer in the hydrogel is undeformed and in mechanical equilibrium:
\begin{equation}
    \mathcal{H} \rightarrow \mathcal{A}, \quad \phi_p \rightarrow \phi_{0p} \mbox{ as } \rho \rightarrow \infty. \label{eq:supbc7}
\end{equation} \label{eq:supallthebc}
\end{subequations}

\subsection{Perturbation Expansion in $\epsilon$}
Working at $\mathcal{O}(\epsilon^{-1})$ in $\epsilon$, (\ref{eq:supgoverning5}) simplifies to give
\begin{equation}
    \frac{\partial^2 \xi^{*}}{\partial {z^{*}}^2} = 0 + \mathcal{O}\left( \epsilon \right), \nonumber
\end{equation}    
with boundary conditions
\begin{equation}
    \xi^{*} \Big{|}_{0} = \frac{\partial \xi^{*}}{\partial z^{*}} \Big{|}_{\mathcal{H}} = 0 \Longrightarrow \xi^{*} = 0 + \mathcal{O}\left( \epsilon \right), \label{eq:supexpansion1}
\end{equation}
(namely $\xi$ and thus $u_p$ are not leading order terms). Similarly, to find $\mathcal{P}$ we expand out (\ref{eq:supgoverning3}) and then apply (\ref{eq:supbc5}), yielding
\begin{equation}
\frac{\partial \mathcal{P}}{\partial z^{*}} = 0 + \mathcal{O}(\epsilon) \Longrightarrow \mathcal{P} = \mathcal{P}_0 + \frac{\partial \zeta^{*}}{\partial z^{*}}\Big{|}_{\mathcal{H}} + \mathcal{O}(\epsilon). \label{eq:supexpansion2}
\end{equation}
Working at $\mathcal{O}(\epsilon^{0})$ in $\epsilon$ and utilising (\ref{eq:supexpansion2}), (\ref{eq:supgoverning6}) simplifies to
\begin{equation}
 \frac{\partial^2 \zeta^{*}}{\partial {z^{*}}^2} = 0 + \mathcal{O}\left( \epsilon \right). \label{eq:supexpansion3} 
\end{equation}
Utilising both (\ref{eq:supgoverning7}) and (\ref{eq:supexpansion1}), the solid phase kinematic boundary condition given in (\ref{eq:supbc2}) simplifies to
\begin{equation}
    \frac{\partial \mathcal{H}}{\partial \tau} \left( 1 - \frac{\partial \zeta^{*}}{\partial z^{*}}\Big{|}_{\mathcal{H}} \right) = \frac{\partial \zeta^{*}}{\partial \tau}\Big{|}_{\mathcal{H}}. \label{eq:supexpansion4a}
\end{equation}
Thus, applying this simplified boundary condition together with (\ref{eq:supbc1}), we integrate (\ref{eq:supexpansion3}) twice to obtain
\begin{align}    
    \zeta^{*} &= z^{*}\left( 1 - \frac{\mathcal{A}}{\mathcal{H}} \right) + \mathcal{O}\left( \epsilon \right), \label{eq:supexpansion4} \\
    w_p^{*} &= \frac{z^{*}}{\mathcal{H}}\frac{\partial \mathcal{H}}{\partial \tau} +  \mathcal{O}\left( \epsilon \right), \label{eq:supexpansion5} \\
    \mathcal{P} &= \mathcal{P}_0 + 1 - \frac{\mathcal{A}}{\mathcal{H}} + \mathcal{O}(\epsilon). \label{eq:supexpansion6}
\end{align}
As a check, note that from (\ref{eq:supexpansion4}), we recover $\zeta^{*}|_{z^*=\mathcal{H}} = \mathcal{H} - \mathcal{A}$. Utilising (\ref{eq:supexpansion5}), the polymer volume fraction mass conservation equation given in (\ref{eq:supgoverning1}) simplifies to give
\begin{equation}
    \frac{\partial}{\partial \tau}\left( \mathcal{H}\phi_p \right) + z^{*} \frac{\partial \mathcal{H}}{\partial \tau} \frac{\partial \phi_p}{\partial z^{*}} = 0 + \mathcal{O}(\epsilon). \label{eq:supexpansion7}
\end{equation}
This admits the separable general solution
\begin{equation}
    \phi_p = \phi_{0p}\int  \tilde{\mathcal{A}} \left( \frac{\mathcal{A}}{\mathcal{H}} \right)^{1 + \sigma} {z^{*}}^{\sigma} d\sigma + \mathcal{O}(\epsilon), \label{eq:supexpansion8}
\end{equation}
where $\tilde{\mathcal{A}} = \tilde{\mathcal{A}}(\rho,\sigma)$ is independent of $\tau$ and $z^{*}$. For mathematical simplicity, we will assume that $\phi_{0p}$ is independent of $z^{*}$ when $\tau=0$; equivalently the dominant mode in (\ref{eq:supexpansion8}) is the $\sigma = 0$ mode giving
\begin{equation}
    \phi_{p} = \frac{\mathcal{A} \phi_{0p}}{\mathcal{H}} + \mathcal{O}\left( \epsilon \right). \label{eq:supexpansion9}
\end{equation}
Finally, integrating (\ref{eq:supgoverning2}) in the $z^{*}$ direction, utilising  (\ref{eq:supgoverning4}) and the boundary conditions given in (\ref{eq:supbc2}), yields the continuity equation for the deflection of the sheet $\overline{\mathcal{H}} = \mathcal{H} - \mathcal{A}$
\begin{align}
    \frac{\partial \overline{\mathcal{H}}}{\partial \tau} &= \frac{1}{\rho}\frac{\partial}{\partial \rho}\left( \rho \int^{\overline{\mathcal{H}} + \mathcal{A}}_0 k \frac{\partial \mathcal{P}}{\partial \rho} dz^{*} \right) \nonumber \\
    &= \frac{\mathcal{A}}{\rho}\frac{\partial}{\partial \rho}\left( \frac{\rho k}{\overline{\mathcal{H}} + \mathcal{A}}\frac{\partial \overline{\mathcal{H}}}{\partial \rho} \right), \label{eq:supexpansion10}
\end{align}
where we have used (\ref{eq:supexpansion9}) to write $k(\phi)$ as a function of $\overline{\mathcal{H}}$.
\section{Self-similar solutions} \label{selfsimilarappendix}
\subsection{Small Deformation Limit}
In the small deformation limit when $\mathrm{max}(\overline{\mathcal{H}}) \ll 1$, the deformation height profile $\overline{\mathcal{H}} = \overline{\mathcal{H}}_s$ satisfies
\begin{subequations}
\begin{equation}
    \frac{\partial \overline{\mathcal{H}}_s}{\partial  \tau} = \nabla^2 \overline{\mathcal{H}}_s, 
\end{equation}   
with volume conservation condition
\begin{equation}    
    \int^{\infty}_0 \rho \overline{\mathcal{H}}_s d\rho = \mathcal{V}.
\end{equation} \label{eq:appendixsmalldeformation}
\end{subequations}
Trying a self-similar solution of the form $f_s(\eta)/{\mathcal{R}}_s^2$ where $\eta = \rho/\mathcal{R}_s$ and $\mathcal{R}_s = \mathcal{R}_s(\tau)$, (\ref{eq:appendixsmalldeformation}) becomes
\begin{equation}
    -\mathcal{R}_s \frac{\partial \mathcal{R}_s}{\partial \tau}\frac{\partial}{\partial \eta}\left( \eta^2 f_s \right) = \frac{\partial}{\partial \eta} \left( \eta \frac{\partial f_s}{\partial \eta} \right).
\end{equation}
Utilising separation of variables and the initial condition $\mathcal{R}_s(t = 0) = 1$, the $\tau$ dependent terms become
\begin{equation}
    \frac{\partial}{\partial \tau}\left( \mathcal{R}_s^2 \right) = \lambda \Longrightarrow \mathcal{R}_s = (1 + \lambda \tau)^{1/2},
\end{equation}
where $\lambda$ is a constant. Similarly, using regularity of $\overline{\mathcal{H}}_s$ at $\eta = 0$ and the initial condition $\overline{\mathcal{H}_s}(\rho = 0, \tau = 0) = 1-\mathcal{A}$, the $\eta$ dependent terms become
\begin{align}
    -\frac{\lambda}{2}\frac{\partial}{\partial \eta}\left( \eta^2 f_s \right) &= \frac{\partial}{\partial \eta}\left( \eta \frac{\partial f_s}{\partial \eta} \right) \nonumber \\
    \Longrightarrow \frac{\partial f_s}{\partial \eta} &= -\frac{\lambda \eta}{2}f_s \nonumber \\
    \Longrightarrow f_s &= (1-\mathcal{A}) \exp{\left( -\frac{\lambda \eta^2}{4} \right)}.
\end{align}
Applying the volume conservation condition in (\ref{eq:appendixsmalldeformation}), $\lambda = 2(1-\mathcal{A})/\mathcal{V}$, we recover the self similar solution given in (9a) of the main text
\begin{equation}
\mathcal{H}_{s} - \mathcal{A} = \frac{1}{R_s^2} \exp{\left(-\frac{(1-\mathcal{A})}{2\mathcal{V}}\left(\frac{\rho}{R_s}\right)^2\right)},
\end{equation}
where
\begin{equation}
\mathcal{R}_s = \left( 1 + \frac{2(1-\mathcal{A}) \tau}{\mathcal{V}} \right)^{1/2}.
\end{equation}
\subsection{Large deformation limit}
Similarly in the large deformation limit, when 
\begin{equation}
\mathrm{max}(\overline{\mathcal{H}}) \gg \mathcal{A}, \nonumber
\end{equation}
the deformation height profile $\overline{\mathcal{H}} = \overline{\mathcal{H}}_l$ satisfies
\begin{subequations}
\begin{equation}
    \frac{\partial \overline{\mathcal{H}}_l}{\partial  \tau} = {\mathcal{A}}^{1 - \beta}\frac{1}{\rho}\frac{\partial}{\partial \rho}\left( \rho {\overline{\mathcal{H}}_l}^{\beta - 1} \frac{\partial \overline{\mathcal{H}}_l}{\partial \rho} \right), 
\end{equation}    
with volume conservation condition
\begin{equation}    
    \int^{\infty}_0 \rho \overline{\mathcal{H}}_l d\rho = \mathcal{V}.
\end{equation} \label{eq:appendixlargedeformation}
\end{subequations}
Trying a self-similar solution of the form $f_l(\eta)/{\mathcal{R}}_l^2$, where $\eta = \rho/\mathcal{R}_l$ and $\mathcal{R}_l = \mathcal{R}_l(\tau)$, (\ref{eq:appendixlargedeformation}) becomes
\begin{equation}
    -\mathcal{R}_l^{2\beta - 1} \frac{\partial \mathcal{R}_l}{\partial \tau}\frac{\partial}{\partial \eta}\left( \eta^2 f_l \right) = {\mathcal{A}}^{1 - \beta}\frac{\partial}{\partial \eta} \left( \eta  f_l^{\beta - 1}\frac{\partial f_l}{\partial \eta} \right).
\end{equation}
Utilising separation of variables and the initial condition $\mathcal{R}_l(t = 0) = 1$, the $\tau$-dependent terms become
\begin{equation}
    \frac{\partial}{\partial \tau}\left( \mathcal{R}_l^{2\beta} \right) = \lambda \Longrightarrow \mathcal{R}_l = (1 + \lambda \tau)^{\frac{1}{2\beta}},
\end{equation}
where $\lambda$ is a constant. Similarly, using regularity of $\overline{\mathcal{H}}_l$ at $\eta = 0$ and the initial condition $\overline{\mathcal{H}}_l(\rho = 0, \tau = 0) = 1-\mathcal{A}$, the $\eta$ dependent terms become
\begin{align}
    &-\frac{\lambda {\mathcal{A}}^{\beta - 1}}{2\beta}\frac{\partial}{\partial \eta}\left( \eta^2 f_l \right) = \frac{\partial}{\partial \eta}\left( \eta f_l^{\beta - 1} \frac{\partial f_l}{\partial \eta} \right) \nonumber \\
    &\Longrightarrow \frac{\partial}{\partial \eta}\left( f_l^{\beta - 1} \right) = -\frac{\lambda \eta (\beta - 1) \mathcal{A}^{\beta - 1}}{2\beta},  \nonumber
\end{align}  
hence
\begin{equation}    
    f_l = (1 - \mathcal{A})\left( 1 - \frac{\lambda (\beta - 1) \mathcal{A}^{\beta - 1}}{4\beta{(1 - \mathcal{A})^{\beta - 1}}}\eta^2 \right) ^{1/(\beta - 1)}.\label{eq:fl}
\end{equation}
Applying the volume conservation condition in (\ref{eq:appendixsmalldeformation}), we find
\begin{equation}
\lambda = \frac{2(1 - \mathcal{A})^{{\beta}}}{\mathcal{V}\mathcal{A}^{\beta - 1}}\sbd{.}
\end{equation}
Thus, we recover the self-similar solution given in equation (9b) of the main text
\begin{align}
&\mathcal{H}_{l} - \mathcal{A} = \nonumber \\
\frac{(1 - \mathcal{A})}{R_l^2} \Bigg{(} 1 - &\frac{{(1 - \mathcal{A})}(\beta - 1)}{2\beta \mathcal{V}}\left( \frac{\rho}{\mathcal{R}_l} \right)^2 \Bigg{)}^{1/(\beta - 1)}, 
\end{align}
where
\begin{equation}
\mathcal{R}_l = \left( 1 + \frac{2 \tau(1 - \mathcal{A})^{{\beta}} {\mathcal{A}}^{1-\beta}}{\mathcal{V}} \right)^{1/2\beta}. \label{eq:rl}
\end{equation}

\begin{figure*}
    \centering
        \includegraphics[width=\textwidth,keepaspectratio=true]{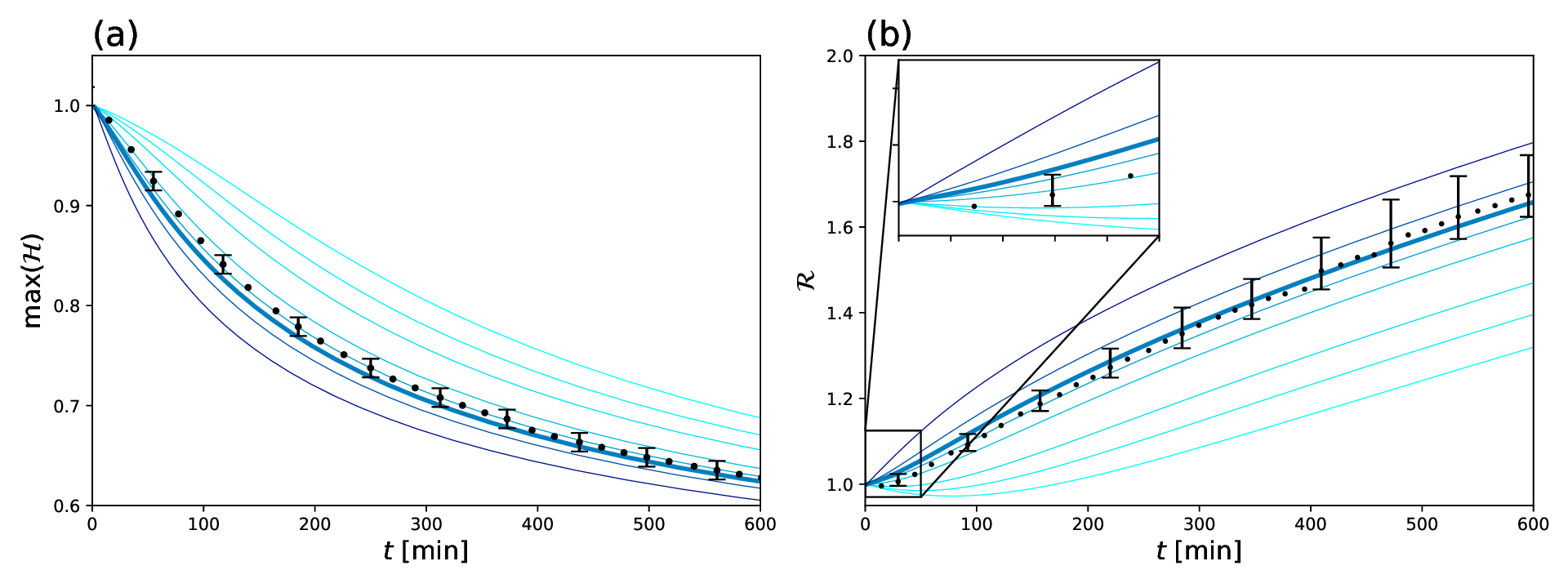}
    \caption{
    The experimental temporal evolution of (a) the scaled maximum blister height $\text{max}(\mathcal{H})$ formed by a \SI{50}{\mu\litre} droplet (dots) and (b) the characteristic blister radius $\mathcal{R}$ superimposed onto a set of blue lines computed using the theoretical model (equation \ref{eq:supexpansion10}), utilising the initial experimental height profile as the initial condition, where $\beta \in \{ 2/3,1,4/3,11/6,25/12,9/4,5/2,3 \}$. Darker blue curves correspond to larger values of $\beta$. The dimensionless time scale $\tau$ was matched to real time $t$ using the relation $\Omega$ = \SI{6.72e9}{\second \per \metre \squared} obtained as best fit for $\beta=2.25$ (thicker blue line). 
    }
    \label{fig:predictedexpbynumerics}
\end{figure*}

\section{Matching Experiment and Model}
\subsection{Numerical Solutions} \label{fenicsappendix}
For general $\mathcal{A}$, (\ref{eq:supexpansion10}) does not admit an analytic solution. Instead, we solve it numerically using the finite-element software package FEniCS (\citep{AlnaesBlechta2015a,LoggMardalEtAl2012a}) on a domain ${\mathcal{Q}:0<\rho<100}$, bounded by no flux boundary conditions. The time derivative is discretised using the backward Euler scheme
\begin{equation}
\frac{\partial \mathcal{H}}{\partial \tau} \approx \frac{(\mathcal{H} - \tilde{\mathcal{H}})}{\delta \tau},
\end{equation}
where $\tilde{\mathcal{H}}$ is taken to be the value of $\mathcal{H}$ at the previous time step. This leads to the weak form of (\ref{eq:supexpansion10})
\begin{align}
    \int  \rho\tilde{\mathcal{H}} Q  d{\rho} &= \mathcal{A}^{1 - \beta} \delta \tau \int \rho (\epsilon_n+ \mathcal{H})^{\beta-1} \frac{\partial \mathcal{H}}{\partial \rho} \frac{\partial Q}{\partial \rho} d{\rho} \nonumber \\
    &+ \int \rho \mathcal{H} Q d{\rho},
\end{align}
where $Q$ is a test function and $\epsilon_n=10^{-6}$ is introduced to avoid NaN errors in the Newton solver. The solutions were calculated with timestep $\delta\tau = \num{e-3}$ and constant grid spacing $\delta \rho = \num{e-2}$. For a given blister height profile $\mathcal{H} = \mathcal{H}(\rho,\tau)$, the effective radius $\mathcal{R} = \mathcal{R}(\mathcal{A},\tau)$ is defined similarly as for the experimental profiles as the point where the deflection of the sheet is half the maximum deflection of the sheet, namely mathematically
\begin{equation}
\mathcal{R} = \left\{ \rho : \mathcal{H}(\rho)-\mathcal{A} = \frac{1}{2}\left(\underset{\tilde{\rho}}{\mathrm{max}}\left(\mathcal{H}(\tilde{\rho})\right)-\mathcal{A} \right) \right\}.\label{eq:practicalmathcalr}
\end{equation}
Our numerical scheme is implicitly validated by the excellent agreement between the similarity solutions and the numerics for small deformation that is shown in Fig. 3 in the main text.

\subsection{Fitting procedure}
We used the numerical solver described above to solve (\ref{eq:supexpansion10}) numerically, utilising an experimentally observed initial condition. To ensure compatibility with the modelling assumptions we chose an experimental profile obtained at $t=t_0$ after the buckling instability was no-longer visible.
Table \ref{tab:initialconditiondata} gives the associated experimental times $t_0$, the corresponding frame number to enable cross referencing to the Electronic Supplementary Material (ESI) and the derived values for each data set for $\{ R_0$, \, $H_0$, \, $\mathcal{A} \}$. 

Since numerical solutions predict the evolution of the blister as a function of dimensionless time $\tau$ with a single fitted parameter $\beta$, the poroelastic time scale  $\tau_0 = \mu_f R_0^2 / \kappa_0 (K + 4G/3)$ for the propagation of the swollen layer through the hydrogel during the initial absorption of the droplet has to be determined for each experiment through fitting. Note that $\tau_0$ can be split into a part that is experiment-dependent and a part that is experiment-independent, namely $\tau_0 = \Omega \, R_0^2 $, where the experiment-dependent part  $R_0^2$ is found through image analysis while the experiment-independent part $\Omega = \mu_f / \kappa_0 (K + 4G/3)$ is an unknown constant fitted to all the experimental data. Furthermore, note that we do not need to fit the elastic moduli $K$ and $G$ individually, decreasing the number of free variables we need to fit from the experimental data and thus increasing the strength of the fit.

In particular, for a given value for $\beta$ we fitted $\Omega$ through a simultaneous fit for the maximum blister height $\text{max}(\mathcal{H})$ to all the experimental data sets, minimising using a least-squares method the objective function
\begin{align}
\sum_i \Bigg{\{}\underset{j}{\mathrm{avg}}\Bigg{(} & \left[{ \text{max}(\mathcal{H}_{e}}(t_j))- \max{\left(\mathcal{H}_\beta\left(\frac{t_j}{\Omega R^2_0}\right)\right)} \right] \nonumber \\
&/ \text{max}(\mathcal{H}_{e}(t_j)) \Bigg{)}^2\Bigg{\}}_i. \label{eq:gelobjectivefunction}
\end{align}
Here, $\Omega$ is the fitted variable, $i$ iterates over all datasets, $j$ iterates over all points within a particular dataset for the experimental blister height profile $\mathcal{H}_e(t_j)$ and $\mathcal{H}_{\beta_1}(\tau_1)$ is the numerical blister height profile at $\tau = \tau_1$ for $\beta = \beta_1$. Both this minimisation and the subsequent linear interpolation to calculate $\mathcal{H}_\beta$ at the experimental time steps were performed using standard NumPy and SciPy methods in Python \citep{2020SciPy-NMeth,harris2020}. 

Fig. \ref{fig:predictedexpbynumerics} explores graphically how varying $\beta$ affects the temporal evolution of the numerical solutions for $\text{max}{\left(\mathcal{H}\right)}$ and $\mathcal{R}$. As a first sweep we calculated the evolution of each profile for $2/3\leq \beta <3$ in increments of 1/3. We then increased the resolution between $\beta=[4/3,7/3]$ until we investigated $\beta=[25/12,26/12,27/12]$. The final values adopted were those obtained for the $\beta$ with the lowest value of the objective function in (\ref{eq:gelobjectivefunction}), namely $\beta = 2.25\pm0.083$ and $\Omega = 6.72 \times 10^{9}\,\si{\second} \, \si{\metre^{-2}}$ which corresponds to poroelastic time scales $\tau_0 = \Omega R_0^2 = 4.14-\SI{36.79}{\hour}$, as shown in table \ref{tab:initialconditiondata}. 

\subsection{Detailed Comparison between Model and Experiment}
In the main text (Fig. 4), we demonstrated excellent agreement between the model and experiments for both $\mathrm{max}(\mathcal{H})$ and full spatial profiles at a fixed late time. In addition, here in the supplementary material in Fig. \ref{fig:radii} we compare the blister radius $\mathcal{R}$ obtained from experimental data with corresponding numerical solution generated using the fitted parameters. We find very good agreement, generally well within the uncertainty range of the experiment for all observation times, noting that the experimental uncertainty associated with $\mathcal{R}$ is much greater than the corresponding uncertainty associated with $\text{max}{\left( \mathcal{H} \right)}$. However, both the smallest \SI{5}{\micro \litre} drop (the darkest blue curve) and the \SI{30}{\micro \litre} drop (the third curve from the bottom) might indicate some systematic deviations, which could be attributed to imperfections in our assumptions (in particular conservation of volume). However, given the good agreement shown in the main text in Fig. 4 and the inset, it is more likely that these are indeed the limits of the experimental setup. The determination of $\mathcal{R}$ becomes more difficult with decreasing slope of the blister (see also (\ref{eq:dr})).

Fig. \ref{fig:profiles} extends the inset of Fig. 4 of the main text by showing the temporal evolution of the full blister height profile for the \SI{50}{\micro \litre} droplet, confirming that the good agreement that is shown in the inset exists for all times.

It is apparent from Fig. \ref{fig:predictedexpbynumerics} that the long-time behaviour of the experiment could be adequately described for models with a wider range of different $\beta$ values by slightly adjusting $\Omega$. This is reflected in that the objective function (\ref{eq:gelobjectivefunction}) only slightly varies for a reasonably wide range of $\beta$. For example, for $2.083\leq\beta\leq 2.583$ (\ref{eq:gelobjectivefunction}) ranges from $\num{4.9e-4}$ to $\num{3e-4}$ (both local maxima) with the minimum (corresponding to $\beta = 2.25$) at \num{2.58e-4}. If we had simply chosen $\beta = 1.833$, the highest value given in the literature, the objective function would have been only \num{5.7e-4} (with corresponding reduced $\Omega=\SI{5.82e9}{\second \per \metre \squared}$). Hence, the agreement with the given data would remain good. However, for smaller values of $\beta$ the objective function increases by an order of magnitude (e.g. for $\beta=2/3$ it becomes \num{6e-3}), demonstrating that the previously proposed value of $\beta=2/3$ \citep{Etzold:2021} is not appropriate for this hydrogel system. 

In conclusion, whilst our experiments demonstrate the importance of the non-linearity in (\ref{eq:supexpansion10}), we find that the exact value of $\beta$ is relatively unimportant if chosen within the correct range. Based on the similarity solutions for large deformation (\ref{eq:appendixlargedeformation}), we expect $\beta$ to have a stronger influence at long observation times during which the radius changes over several orders of magnitude. However, this was not possible within the constraints of the current experimental setup.
\begin{table}[h]
    \centering
        \caption{Experimental $t_0$ from which the initial condition for the numerical solution was obtained and corresponding values $\{ H_0, \, R_0, \, \mathcal{A}\}$. The second column refers to the frame number in the corresponding raw dataset provided as ESI.}
    \begin{tabular}{|c|c|c|c|c|c|c|}
        \hline
        $V [\si{\micro \litre}]$ & Frame & $t_0 [\si{\minute}]$ & $H_0$ [\si{\milli \metre}]  & $R_0$ [\si{\milli \metre}] & $\mathcal{A}$ & $\tau_0$ [\si{\hour}]\\
        &&&&&&\\
        \hline
        100 & 180 & 82 & 2.61 & 4.44 & 0.43 & 36.79\\
        50 & 220 & 39.5 & 2.44 & 3.27 & 0.46& 19.96 \\
        30 & 782 &21.0 & 2.28 & 2.78 & 0.49& 14.42\\
        25 & 235 &26.0 & 2.18 & 2.61 & 0.51 & 12.71\\
        10 & 109 &20.0 & 1.82 & 1.94 & 0.62 & 7.02\\
        5 & 358 &9.5 & 1.66 & 1.49 & 0.68 & 4.14 \\
        \hline
    \end{tabular}

    \label{tab:initialconditiondata}
\end{table}

\begin{figure}[!h]
    \centering
    \includegraphics[width = \columnwidth]{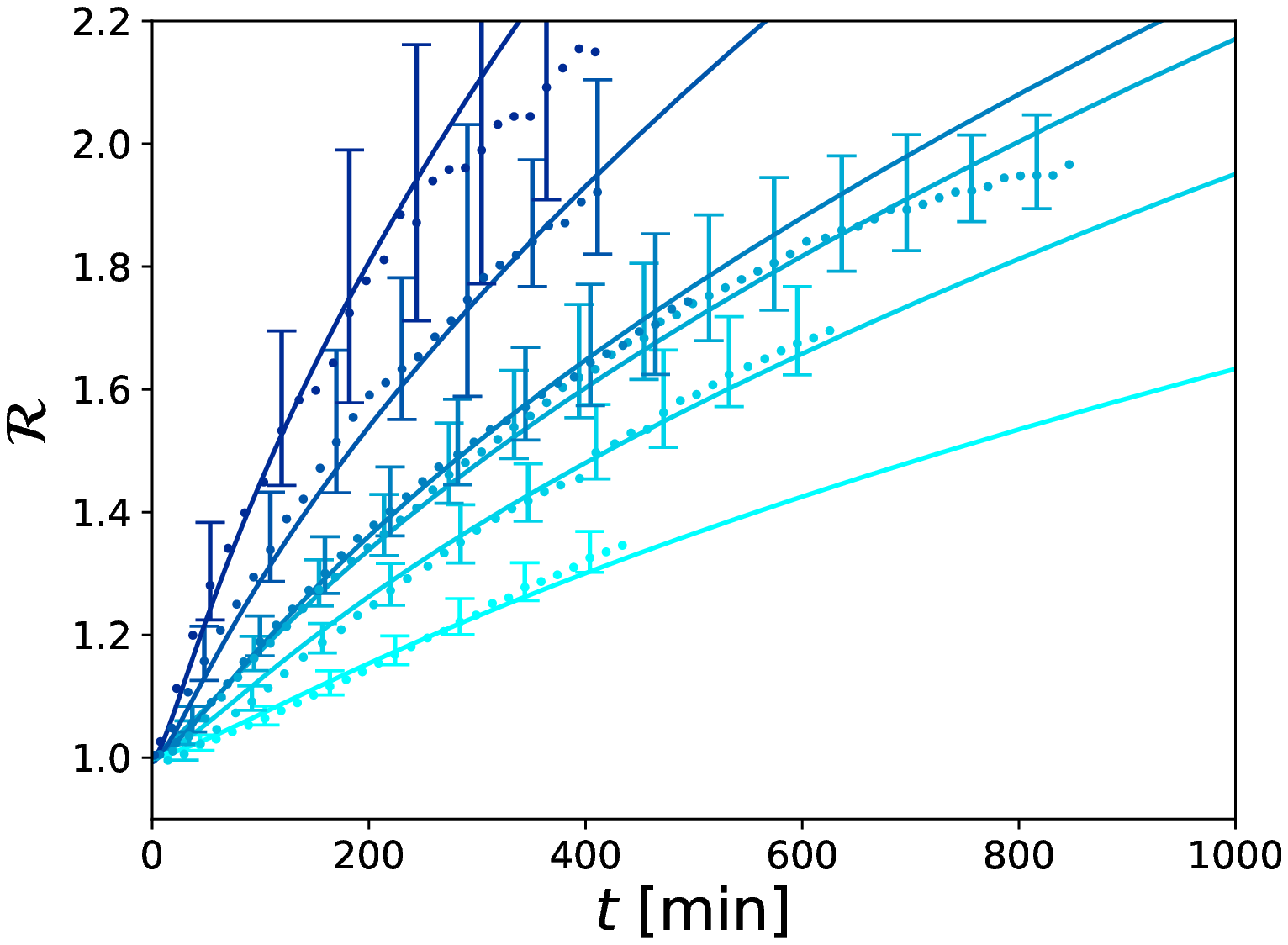}
        \caption{The temporal evolution of the scaled blister radius $\mathcal{R}$ for a range of initial droplet volumes, plotting experimental data with circles and numerical solutions with lines (solving (\ref{eq:supexpansion10}) utilising the fitted values for $\Omega$ and $\beta$). Here, darker colours denote smaller droplets. Numerical solutions are generated using initial conditions taken directly from the experimental data that they are fitted against.}
    \label{fig:radii}
\end{figure}

\begin{figure}[!h]
    \centering
    \includegraphics[width=0.5\textwidth,keepaspectratio=true]{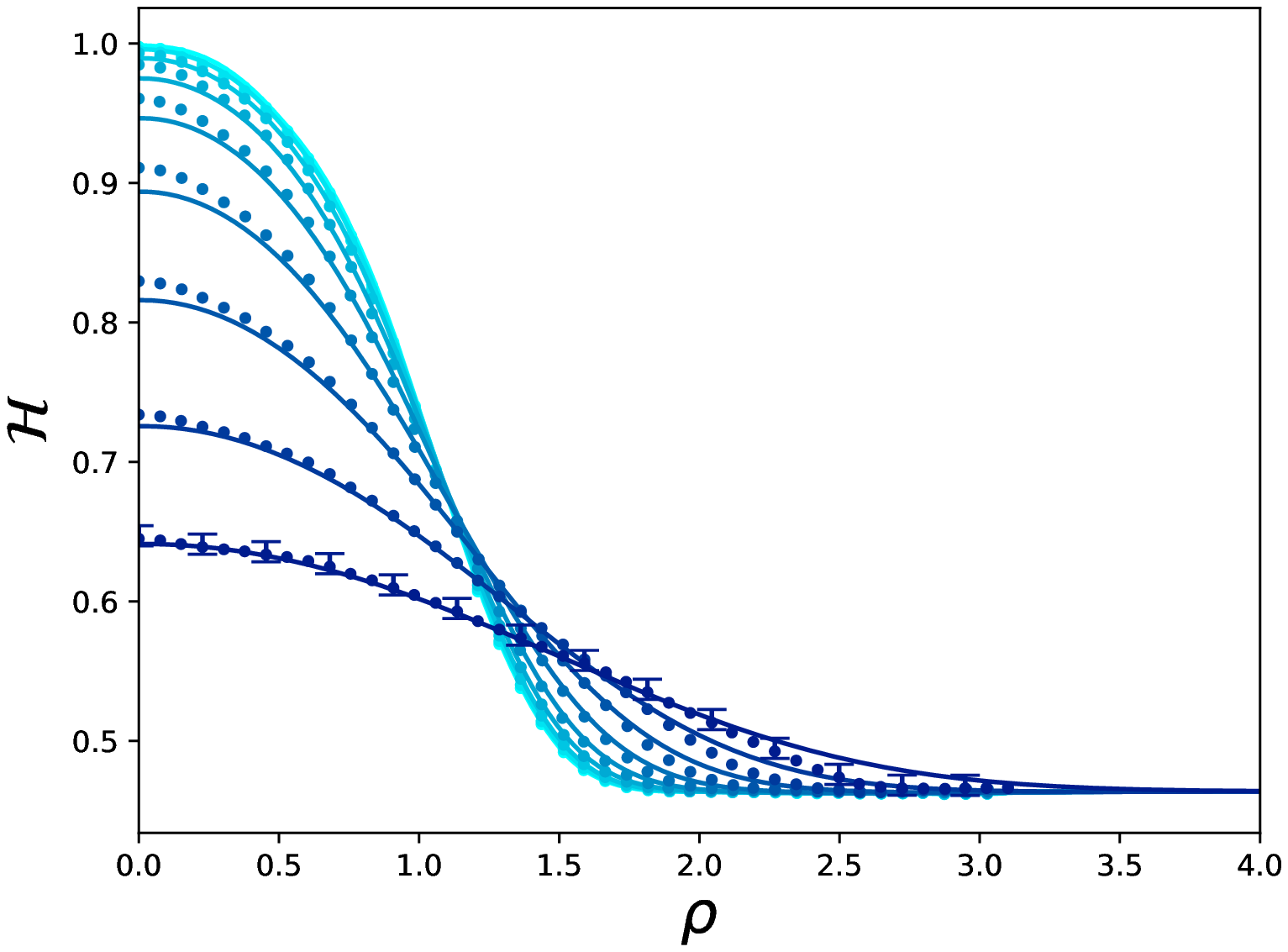}
    \caption{Comparison between experimentally observed swelling and numerical solutions of the poroelastic model, plotting the spatial dependence of the scaled height $\mathcal{H}$ for a blister formed by a \SI{50}{\mu\litre} droplet at a range of different times during the experiment. Here, experimental data is denoted by circles while numerical solutions are denoted by lines. The lightest blue curve corresponds to $t = \SI{2}{\minute}$ with each subsequent darker blue curve twice the previous time. Numerical solutions are generated solving (\ref{eq:supexpansion10}) using initial conditions taken directly from the experimental data that they are fitted against.}
    \label{fig:profiles}
\end{figure}

\bibliography{hydrogel_references}